\documentclass[11pt]{article} 

\usepackage{textcomp, amssymb}  
\usepackage{anysize}            
\usepackage{amsfonts}
\usepackage{amsmath}
\usepackage{dsfont}
\usepackage{MnSymbol}
\usepackage{mathtools}
\usepackage{graphicx}
\usepackage{epsfig}
\usepackage{epstopdf}
\usepackage{tikz}
\usepackage{amsthm}
\usepackage{ifpdf}
\usepackage{thm-restate}
\usepackage{stmaryrd}
\usepackage{cleveref}
\usepackage{subcaption}

\usetikzlibrary{calc,decorations.pathmorphing,shapes,graphs,positioning,automata,angles}

\newcounter{sarrow}

\newcounter{sarrow1}
\newcommand\xnrsquigarrow[1]{%
\stepcounter{sarrow1}%
\mathrel{\begin{tikzpicture}[baseline= {( $ (current bounding box.south) + (0,-0.5ex) $ )}]
\node[inner sep=.5ex] (\thesarrow) {$\scriptstyle #1$};
\path[draw,<-,decorate,
  decoration={zigzag,amplitude=0.7pt,segment length=1.2mm,pre=lineto,pre length=4pt}]
    (\thesarrow1.south east) -- (\thesarrow1.south west);
    $\slashedarrowfill@\relbar\relbar/$
    \end{tikzpicture}}%
}

\makeatletter
\def\slashedarrowfill@#1#2#3#4#5{%
  $\m@th\thickmuskip0mu\medmuskip\thickmuskip\thinmuskip\thickmuskip
   \relax#5#1\mkern-7mu%
   \cleaders\hbox{$#5\mkern-2mu#2\mkern-2mu$}\hfill
   \mathclap{#3}\mathclap{#2}%
   \cleaders\hbox{$#5\mkern-2mu#2\mkern-2mu$}\hfill
   \mkern-7mu#4$%
}
\def\rightslashedarrowfillb@{%
  \slashedarrowfill@\relbar\relbar/\rightarrow}
\newcommand\xnrightarrow[2][]{%
  \ext@arrow 0055{\rightslashedarrowfillb@}{#1}{#2}}

\def\rightslashedarrowfille@{%
  \slashedarrowfill@\relbar\relbar/\twoheadrightarrow}
\newcommand\xntworightarrow[2][]{%
  \ext@arrow 0055{\rightslashedarrowfille@}{#1}{#2}}

\def\rightslashedarrowfillg@{%
  \slashedarrowfill@\relbar\relbar{\raisebox{.12em}{}}\twoheadrightarrow}
\newcommand\xtworightarrow[2][]{%
  \ext@arrow 0055{\rightslashedarrowfillg@}{#1}{#2}}

\def\rightslashedarrowfillx@{%
  \slashedarrowfill@\Relbar\Relbar/\rightrightarrows}
\newcommand\xnTworightarrow[2][]{%
  \ext@arrow 0055{\rightslashedarrowfillx@}{#1}{#2}}

\def\rightslashedarrowfilly@{%
  \slashedarrowfill@\Relbar\Relbar{\raisebox{.12em}{}}\rightrightarrows}
\newcommand\xTworightarrow[2][]{%
  \ext@arrow 0055{\rightslashedarrowfilly@}{#1}{#2}}

\pgfdeclareshape{slash underlined}
{
  \inheritsavedanchors[from=rectangle] 
  \inheritanchorborder[from=rectangle]
  \inheritanchor[from=rectangle]{north}
  \inheritanchor[from=rectangle]{north west}
  \inheritanchor[from=rectangle]{north east}
  \inheritanchor[from=rectangle]{center}
  \inheritanchor[from=rectangle]{west}
  \inheritanchor[from=rectangle]{east}
  \inheritanchor[from=rectangle]{mid}
  \inheritanchor[from=rectangle]{mid west}
  \inheritanchor[from=rectangle]{mid east}
  \inheritanchor[from=rectangle]{base}
  \inheritanchor[from=rectangle]{base west}
  \inheritanchor[from=rectangle]{base east}
  \inheritanchor[from=rectangle]{south}
  \inheritanchor[from=rectangle]{south west}
  \inheritanchor[from=rectangle]{south east}
  \inheritanchorborder[from=rectangle]
  \foregroundpath{
    \southwest \pgf@xa=\pgf@x \pgf@ya=\pgf@y
    \northeast \pgf@xb=\pgf@x \pgf@yb=\pgf@y
    \pgf@xc=\pgf@xa
    \advance\pgf@xc by .5\pgf@xb
    \pgf@yc=\pgf@ya 
    \advance\pgf@xc by -1.3pt
    \advance\pgf@yc by -1.8pt
    \pgfpathmoveto{\pgfqpoint{\pgf@xc}{\pgf@yc}}
    \advance\pgf@xc by  2.6pt
    \advance\pgf@yc by  3.6pt
    \pgfpathlineto{\pgfqpoint{\pgf@xc}{\pgf@yc}}
    \pgfpathmoveto{\pgfqpoint{\pgf@xa}{\pgf@ya}}
    \pgfpathlineto{\pgfqpoint{\pgf@xb}{\pgf@ya}}
 }
}
\tikzset{nomorepostaction/.code=\let\tikz@postactions\pgfutil@empty}

\newcommand{\semangle}[1]{\langle\!|#1|\!\rangle}
\newcommand{\sembrack}[1]{\llbracket #1\rrbracket}
\newcommand*{\rmbrace}{|\mskip-4mu\}}
\newcommand*{\lmbrace}{\{\mskip-4mu|}
\newcommand*{\mset}[1]{\lmbrace#1\rmbrace}

\newtheorem{theorem}{Theorem}[section]
\newtheorem{definition}[theorem]{Definition}

\newtheorem{lemma}[theorem]{Lemma}

\newtheorem{corollary}[theorem]{Corollary}

\begin{document}

\begin{titlepage}
\thispagestyle{empty}

\hrule
\begin{center}
{\bf\LARGE Step Automata\\}
%
\vspace{0.5cm}
--- Yong Wang ---

\vspace{2cm}

\end{center}
\end{titlepage}

\newpage 

\setcounter{page}{1}\pagenumbering{roman}

\tableofcontents

\newpage
\setcounter{page}{1}\pagenumbering{arabic}

        \section{Introduction}\label{intro}

For computation, there existed Turing machine \cite{TM1} \cite{TM2} and later-matured automata theory. For low-level parallel computation, there existed variants of Turing machine, such as two-tapes Turing machine \cite{TM6} and multi-tapes Turing machine \cite{TM7}. In the literature, the combination of computation and concurrency is still active, such the combination of automata and processes \cite{AP1} \cite{AP2} \cite{AP3} \cite{AP4} \cite{AP5} \cite{AP6} \cite{AP8}, and the introduction of concurrency into automata: the so-called pomset automata \cite{PA1} \cite{PA2} \cite{PA3} \cite{PA4} and branch automata \cite{BA1} \cite{BA2} \cite{BA3} \cite{BA4}. 

But the linkage of Turing machine and concurrent automaton is still absent. In this paper, we propose the concepts of step automaton and step Turing machine (STM), which a natural extension to traditional automaton and classical Turing machine just allowing an automaton or Turing machine to execute a step of atomic actions (without partial orders pairwise).

As a natural parallel computational model, in future, we believe that an STM can be used in two scenarios at least:

\begin{enumerate}
  \item It can be used to analyze the parallel complexity of a computation. Currently, the main tool of parallel complexity analysis tool is circuit complexity. Since a circuit can be modelled by an STM naturally, we believe that the parallel complexity results by circuits can be re-obtained by STMs. Maybe there exists further results, we do not know. 
  \item The computability of neural networks, generally, recurrent neural networks (RNN) were thoroughly explored \cite{NNC} based on different kinds of weights and different precisions. The emergency of transformers \cite{AIAYN} provided parallelism to make large scale training possible. The analyses of computability of transformers should consider their parallelism nature \cite{PT1} \cite{PT2}, besides its Turing completeness based infinite precision \cite{AIC}. The parallelism nature of a transformer and the parallelism nature of an STM should be matched, i.e., STMs provide a new tool to analyze the computability of transformers, and let them be future works.
\end{enumerate}
\newpage\section{Preliminaries}\label{pre} 

For self-satisfactory, in this section, we introduce the preliminaries on set and language in \cref{sl} and automata in \cref{auto}.

\subsection{Set and Language}\label{sl}

\begin{definition}[Set]
A set contains some objects, and let $\{-\}$ denote the contents of a set. For instance, $\mathbb{N}=\{1,2,3,\cdots\}$. Let $a\in A$ denote that $a$ is an element of the set $A$ and $a\notin A$ denote that $a$ is not an element of the set $A$. For all $a\in A$, if we can get $a\in B$, then we say that $A$ is a subset of $B$ denoted $A\subseteq B$. If $A\subseteq B$ and $B\subseteq A$, then $A=B$. We can define a new set by use of predicates on the existing sets, such that $\{n\in\mathbb{N}|\exists k\in\mathbb{N},n=2k\}$ for the set of even numbers. We can also specify a set to be the smallest set satisfy some inductive inference rules, for instance, we specify the set of even numbers $A$ satisfying the following rules:

$$\frac{}{0\in A}\quad\quad\frac{n\in A}{n+2\in A}$$
\end{definition}

\begin{definition}[Set composition]
The union of two sets $A$ and $B$, is denoted by $A\cup B=\{a|a\in A\textrm{ or }a\in B\}$, and the intersection of $A$ and $B$ by $A\cap B=\{a|a\in A\textrm{ and }a\in B\}$, the difference of $A$ and $B$ by $A\setminus B=\{a|a\in A\textrm{ and }a\notin B\}$. The empty set $\emptyset$ contains nothing. The set of all subsets of a set $A$ is called the powerset of $A$ denoted $2^A$.
\end{definition}

\begin{definition}[Tuple]
A tuple is a finite and ordered list of objects and denoted $\langle -\rangle$. For sets $A$ and $B$, the Cartesian product of $A$ and $B$ is denoted by $A\times B=\{\langle a,b\rangle|a\in A,b\in B\}$. $A^n$ is the $n$-fold Cartesian product of set $A$, for instance, $A^2=A\times A$. Tuples can be flattened as $A\times (B\times C)=(A\times B)\times C=A\times B\times C$ for sets $A$, $B$ and $C$.
\end{definition}

\begin{definition}[Relation]
A relation $R$ between sets $A$ and $B$ is a subset of $A\times B$, i.e., $R\subseteq A\times B$. We say that $R$ is a relation on set $A$ if $R$ is a relation between $A$ and itself, and,

\begin{itemize}
  \item $R$ is reflexive if for all $a\in A$, $aRa$ holds; it is irreflexive if for all $a\in A$, $aRa$ does not hold.
  \item $R$ is symmetric if for all $a,a'\in A$ with $aRa'$, then $a'Ra$ holds; it is antisymmetric if for all $a,a'\in A$ with $aRa'$ and $a'Ra$, then $a=a'$.
  \item $R$ is transitive if for all $a,a',a''\in A$ with $aRa'$ and $a'Ra''$, then $aRa''$ holds.
\end{itemize}
\end{definition}

\begin{definition}[Preorder, partial order, strict order]
If a relation $R$ is reflexive and transitive, we call that it is a preorder; When it is a preorder and antisymmetric, it is called a partial order, and a partially ordered set (poset) is a pair $\langle A,R\rangle$ with a set $A$ and a partial order $R$ on $A$; When it is irreflexive and transitive, it is called a strict order.
\end{definition}

\begin{definition}[Equivalence]
A relation $R$ is called an equivalence, if it is reflexive, symmetric and transitive. For an equivalent relation $R$ and a set $A$, $[a]_R=\{a'\in A|aRa'\}$ is called the equivalence class of $a\in A$.
\end{definition}

\begin{definition}[Relation composition]
For sets $A$, $B$ and $C$, and relations $R\subseteq A\times B$ and $R'\subseteq B\times C$, the relational composition denoted $R\circ R'$, is defined as the smallest relation $a(R\circ R')c$ satisfying $aRb$ and $bR'c$ with $a\in A$, $b\in B$ and $c\in C$. For a relation $R$ on set $A$, we denote $R^*$ for the reflexive and transitive closure of $R$, which is the least reflexive and transitive relation on $A$ that contains $R$.
\end{definition}

\begin{definition}[Function]
A function $f:A\rightarrow B$ from sets $A$ to $B$ is a relation between $A$ and $B$, i.e., for every $a\in A$, there exists one $b=f(a)\in B$, where $A$ is called the domain of $f$ and $B$ the codomain of $f$. $\sembrack{-}$ is also used as a function with $-$ a placeholder, i.e., $\sembrack{x}$ is the value of $\sembrack{-}$ for input $x$. A function $f$ is a bijection if for every $b\in B$, there exists exactly one $a\in A$ such that $b=f(a)$. For functions $f:A\rightarrow B$ and $g:B\rightarrow C$, the functional composition of $f$ and $g$ denoted $g\circ f$ such that $(g\circ f)(a)=g(f(a))$ for $a\in A$.
\end{definition}

\begin{definition}[Poset morphism]
For posets $\langle A,\leq\rangle$ and $\langle A',\leq'\rangle$ and function $f:A\rightarrow A'$, $f$ is called a poset morphism if for $a_0,a_1\in A$ with $a_0\leq a_1$, then $f(a_0)\leq' f(a_1)$ holds.
\end{definition}

\begin{definition}[Multiset]
A multiset is a kind of set of objects which may be repetitive denoted $\mset{-}$, such that $\mset{0,1,1}$ is significantly distinguishable from $\mset{0,1}$.
\end{definition}

\begin{definition}[Alphabet, word, language]
An alphabet $\Sigma$ is a (maybe infinite) set of symbols. A word over some alphabet $\Sigma$ is a finite sequence of symbols from $\Sigma$. Words can be concatenated horizontally and the horizontal concatenation operator is denoted by $\cdot$, for instance $ab\cdot c=abc$. The empty word is denoted $1$ with $1\cdot w=w=w\cdot 1$ for word $w$. For $n\in\mathbb{N}$ and $a\in\Sigma$, $a^n$ is the $n$-fold concatenation of $a$ with $a^0=1$ and $a^{n+1}=a\cdot a^n$. Words can be concatenated vertically and the vertical concatenation operator is denoted by $\parallel$, for instance $a\parallel b\parallel c$. $1\parallel w=w=w\parallel 1$ for word $w$. For $n\in\mathbb{N}$ and $a\in\Sigma$, $a^{\langle n\rangle}$ is the $n$-fold vertical concatenation of $a$ with $a^{\langle 0\rangle}=1$ and $a^{\langle n+1\rangle}=a\parallel a^{\langle n\rangle}$. A language is a set of words, and the language of all words over an alphabet $\Sigma$ is denoted $\Sigma^{\langle *\rangle^*}$.
\end{definition}

\begin{definition}[Expressions]
Expressions are builded by function symbols and constants over a fixed alphabet inductively. For instance, the set of numerical expressions over some fixed set of variables $V$ are defined as the smallest set $\mathcal{T}$ satisfying the following inference rules:

$$\frac{n\in\mathbb{N}}{n\in \mathcal{T}}\quad \frac{v\in V}{v\in \mathcal{T}}\quad \frac{x,y\in \mathcal{T}}{x+y\in \mathcal{T}}\quad \frac{x,y\in \mathcal{T}}{x\times y\in \mathcal{T}}\quad \frac{x\in \mathcal{T}}{-x\in \mathcal{T}}$$

The above inference rules are equal to the following Backus-Naur Form (BNF) grammer.

$$\mathcal{T}\ni x,y::=n\in\mathbb{N}|v\in V|x+y|x\times y|-x$$
\end{definition}

\begin{definition}[Congruence, precongruence]
A relation $R$ on a set of expressions is a congruence if it is an equivalence compatible with the operators; and a relation $R$ on a set of expressions is a precongruence if it is a preorder compatible with the operators.
\end{definition}

\subsection{Automata}\label{auto}

\begin{definition}[Automaton]
An automaton is a tuple $A=(Q,F,\delta)$ where $Q$ is a finite set of states, $F\subseteq Q$ is the set of final states, and $\delta$ is the finite set of the transitions of $A$ and $\delta\subseteq (Q\setminus F)\times \Sigma\times Q$.
\end{definition}

It is well-known that automata recognize rational languages.

\begin{definition}[Transition relation]
Let $p,q\in Q$. We define the transition relation $\xrightarrow[A]{}\subseteq Q\times \Sigma\times Q$ on $A$ as the smallest relation satisfying:

\begin{enumerate}
  \item $p\xrightarrow[A]{1}p$ for all $p\in Q$;
  \item $p\xrightarrow[A]{a}q$ if and only if $(p,a,q)\in \delta$.
\end{enumerate}
\end{definition} 
\newpage\section{Series-Parallel Rational Language and Its Algebra}\label{spria}

In this section, we introduce series-parallel rational language and it algebra modulo language equivalence called Concurrent Kleene Algebra (CKA). We introduce the concepts of pomset and step in \cref{pas}. In \cref{sprl}, we introduce series-parallel rational language, CKA in \cref{tlmle} and series-parallel rational systems in \cref{sprl}.

\subsection{Pomset and Step}\label{pas}

\begin{definition}[Labelled poset]
A labelled poset is a tuple $\mathbf{u}=\langle S, \leq, \lambda\rangle$, where $S$ is the carrier set, $\leq$ is a partial order on $S$ and $\lambda$ is a labelling function $\lambda:S\rightarrow\Sigma$.

For a labelled poset $\mathbf{u}$, $S_{\mathbf{u}}$, $\leq_{\mathbf{u}}$ and $\lambda_{\mathbf{u}}$ denote the carrier, the partial order and the labelling of $\mathbf{u}$ respectively. The set of labelled posets is denoted $\mathsf{LP}$ and the empty labelled poset is $\mathbf{1}$.
\end{definition}

\begin{definition}[Labelled poset isomorphism]
Let $\mathbf{u}=\langle S_1,\leq_1,\lambda_1\rangle$ and $\mathbf{v}=\langle S_2,\leq_2,\lambda_2\rangle$ be labelled posets. A labelled poset morphism $h$ from $\mathbf{u}=\langle S_1,\leq_1,\lambda_1\rangle$ to $\mathbf{v}=\langle S_2,\leq_2,\lambda_2\rangle$ is a poset morphism from $\langle S_1,\leq_1\rangle$ and $\langle S_2,\leq_2\rangle$ with $\lambda_2\circ h=\lambda_1$. Moreover, $h$ is a labelled poset isomorphism if it is a bijection with $h^{-1}$ is a poset isomorphism from $\langle S_2,\leq_2,\lambda_2\rangle$ to $\langle S_1,\leq_1,\lambda_1\rangle$. We say that $\mathbf{u}=\langle S_1,\leq_1,\lambda_1\rangle$ is isomorphic to $\mathbf{v}=\langle S_2,\leq_2,\lambda_2\rangle$ denoted $\langle S_1,\leq_1,\lambda_1\rangle\sim\langle S_2,\leq_2,\lambda_2\rangle$, if there exists a poset isomorphism $h$ between $\langle S_1,\leq_1,\lambda_1\rangle$ and $\langle S_2,\leq_2,\lambda_2\rangle$.
\end{definition}

It is easy to see that $\sim$ is an equivalence and can be used to abstract from the carriers.

\begin{definition}[Pomset and step]
A partially ordered multiset, pomset, is a $\sim$-equivalence class of labelled posets. The $\sim$-equivalence class of $\mathbf{u}\in\mathsf{LP}$ is denoted $[\mathbf{u}]$; the set of pomsets is denoted $\mathsf{Pom}$; the empty labelled poset is denoted $\mathbf{1}$ and the $\sim$-equivalence class of $\mathbf{1}$ is denoted by $1$; the pomset containing exactly one action $a\in\Sigma$ is called primitive. While the pomset containing actions which are pairwise without partial orders , i.e., $\leq_{\mathbf{u}}=\emptyset$, is called a step. Usually, we denote a pomset $U\in\mathsf{Pom}$ as $\mset{a_1,\cdots,a_n}$ and a step as $\semangle{a_1,\cdots,a_n}$ for $a_1,\cdots,a_n\in\Sigma$ and $n\in\mathbb{N}$.
\end{definition}

\begin{definition}[Pomset composition in parallel]
Let $U,V\in\mathsf{Pom}$ with $U=[\mathbf{u}]$ and $V=[\mathbf{v}]$. We write $U\parallel V$ for the parallel composition of $U$ and $V$, which is the pomset represented by $\mathbf{u}\parallel\mathbf{v}$, where

$$S_{\mathbf{u}\parallel\mathbf{v}}=S_{\mathbf{u}}\cup S_{\mathbf{v}}
\quad\quad\leq_{\mathbf{u}\parallel\mathbf{v}}=\leq_{\mathbf{u}}\cup\leq_{\mathbf{v}}
\quad\quad\lambda_{\mathbf{u}\parallel\mathbf{v}}(x)=\begin{cases}
\lambda_{\mathbf{u}}\quad x\in S_{\mathbf{u}}\\
\lambda_{\mathbf{v}}\quad x\in S_{\mathbf{v}}
\end{cases}$$
\end{definition}

\begin{definition}[Pomsetc composition in sequence]
Let $U,V\in\mathsf{Pom}$ with $U=[\mathbf{u}]$ and $V=[\mathbf{v}]$. We write $U\cdot V$ for the sequential composition of $U$ and $V$, which is the pomset represented by $\mathbf{u}\cdot\mathbf{v}$, where

$$S_{\mathbf{u}\cdot\mathbf{v}}=S_{\mathbf{u}}\cup S_{\mathbf{v}}
\quad\quad\leq_{\mathbf{u}\cdot\mathbf{v}}=\leq_{\mathbf{u}}\cup\leq_{\mathbf{v}}
\quad\quad\lambda_{\mathbf{u}\cdot\mathbf{v}}(x)=\begin{cases}
\lambda_{\mathbf{u}}\quad x\in S_{\mathbf{u}}\\
\lambda_{\mathbf{v}}\quad x\in S_{\mathbf{v}}
\end{cases}$$
\end{definition}

The following definitions and conclusions are coming from \cite{CKA7}, we retype them.

\begin{definition}[Pomset types]
Let $U\in\mathsf{Pom}$, $U$ is sequential (resp. parallel) if there exist non-empty pomsets $U_1$ and $U_2$ such that $U=U_1\cdot U_2$ (resp. $U=U_1\parallel U_2$).
\end{definition}

\begin{definition}[Factorization]
Let $U\in\mathsf{Pom}$. (1) When $U=U_1\cdot\cdots \cdot U_i\cdot\cdots \cdot U_n$ with each $U_i$ non-sequential and non-empty, the sequence $U_1,\cdots,U_i,\cdots,U_n$ is called a sequential factorization of $U$. (2) When $U=U_1\parallel\cdots \parallel U_i\parallel\cdots \parallel U_n$ with each $U_i$ non-parallel and non-empty, the multiset $\mset{U_1,\cdots,U_i,\cdots,U_n}$ is called a parallel factorization of $U$.
\end{definition}

\begin{lemma}[Factorization]\label{LemmaFactorization}
Sequential and parallel factorizations exist uniquely.
\end{lemma}

\subsection{Series-Parallel Rational Language}\label{sprl}

\begin{definition}[Series-parallel pomset]
The set of series-parallel pomset, or sp-pomsets denoted $\mathsf{SP}$, is the smallest set satisfying the following rules:

$$\frac{}{1\in\mathsf{SP}}
\quad\frac{a\in\Sigma}{a\in\mathsf{SP}}
\quad\frac{U,V\in\mathsf{SP}}{U\cdot V\in\mathsf{SP}}
\quad\frac{U,V\in\mathsf{SP}}{U\parallel V\in\mathsf{SP}}$$
\end{definition}

\begin{definition}[N-shape]\label{nshapes}
Let $U=[\mathbf{u}]$ be a pomset. An N-shape in $U$ is a quadruple $u_0,u_1,u_2,u_3\in S_{\mathbf{u}}$ of distinct points such that $u_0\leq_{\mathbf{u}}u_1$, $u_2\leq_{\mathbf{u}}u_3$ and $u_0\leq_{\mathbf{u}}u_3$ and their exists no other relations among them. A pomset $U$ is N-free if it has no N-shapes.
\end{definition}

\begin{theorem}[N-shape]
A pomset is series-parallel if and only if it is N-shape-free in Definition \ref{nshape1s}.
\end{theorem}

\begin{definition}[Pomset language]
A pomset language is a set of pomsets. A pomset language made up of sp-pomsetcs is referred to as series-communication-parallel language, or sp-language for short.
\end{definition}

\begin{definition}[Pomset language composition]
Let $L,K\subseteq\mathsf{Pom}$. Then we define the following compositions.

$$L+K=L\cup K\quad L\cdot K=\{U\cdot V:U\in L,V\in K\}\quad L\parallel K=\{U\parallel V:U\in L,V\in K\}$$
$$L^*=\bigcup_{n\in\mathbb{N}}L^n\textrm{ where }L^0=\{1\}\textrm{ and }L^{n+1}=L^n\cdot L$$
$$L^{\langle *\rangle}=\bigcup_{n\in\mathbb{N}}L^{\langle n\rangle}\textrm{ where }L^{\langle 0\rangle}=\{1\}\textrm{ and }L^{\langle{n+1}\rangle}=L^{\langle n\rangle}\parallel L$$
\end{definition}

\begin{definition}[Pomset language substitution]
Let $\Delta$ be an alphabet. A substitution is a function $\zeta:\Sigma\rightarrow 2^{\mathsf{Pom}(\Delta)}$ and lift to $\zeta:\mathsf{Pom}(\Delta)\rightarrow 2^{\mathsf{Pom}(\Delta)}$:

$$\zeta(1)=\{1\}\quad \zeta(U\cdot V)=\zeta(U)\cdot\zeta(V)\quad\zeta(U\parallel V)=\zeta(U)\parallel\zeta(V)$$
\end{definition} 

We define the syntax and language semantics of the series-parallel rational (spr-) expressions.

\begin{definition}[Syntax of spr-expressions]
We define the set of spr-expressions $\mathcal{T}_{SPR}$ as follows.

$$\mathcal{T}_{SPR}\ni x,y::=0|1|a,b\in\Sigma|x+y|x\cdot y|x^*|x\parallel y|x^{\langle *\rangle}$$
\end{definition}

In the definition of spr-expressions, the atomic actions include actions in $a,b\in\Sigma$, the constant $0$ denoted inaction without any behaviour, the constant $1$ denoted empty action which terminates immediately and successfully. The operator $+$ is the alternative composition, i.e., the program $x+y$ either executes $x$ or $y$ alternatively. The operator $\cdot$ is the sequential composition, i.e., the program $x\cdot y$ firstly executes $x$ followed $y$. The Kleene star $x^*$ can execute $x$ for some number of times sequentially (maybe zero). The operator $\parallel$ is the parallel composition, i.e., the program $x\parallel y$ executes $x$ and $y$ in parallel. The parallel star $x^{\langle *\rangle}$ can execute $x$ for some number of times in parallel (maybe zero). 

\begin{definition}[Language semantics of spr-expressions]
We define the interpretation of spr-expressions $\sembrack{-}_{SPR}:\mathcal{T}_{SPR}\rightarrow 2^{\mathsf{SCP}}$ inductively as Table \ref{LSCKA} shows.
\end{definition}

\begin{center}
    \begin{table}
        $$\sembrack{0}_{SPR}=\emptyset \quad \sembrack{a}_{SPR}=\{a\} \quad \sembrack{x\cdot y}_{SPR}=\sembrack{x}_{SPR}\cdot \sembrack{y}_{SPR}$$
        $$\sembrack{1}_{SPR}=\{1\} \quad \sembrack{x+y}_{SPR}=\sembrack{x}_{SPR}+\sembrack{y}_{SPR} \quad\sembrack{x^*}_{SPR}=\sembrack{x}^*_{SPR}$$
        $$\sembrack{x\parallel y}_{SPR}=\sembrack{x}_{SPR}\parallel\sembrack{y}_{SPR} \quad\sembrack{x^{\langle *\rangle}}_{SPR}=\sembrack{x}^{\langle *\rangle}_{SPR}$$
        \caption{Language semantics of spr-expressions}
        \label{LSCKA}
    \end{table}
\end{center}

\subsection{The Algebra Modulo Language Equivalence}\label{tlmle}

We define a concurrent Kleene algebra (CKA) as a tuple $(\Sigma,+,\cdot,^*,\parallel,^{\langle *\rangle},0,1)$, where $\Sigma$ is an alphabet, $^*$ and $^{\langle *\rangle}$ are unary, $+$, $\cdot$ and $\parallel$ are binary operators, and $0$ and $1$ are constants, which satisfies the axioms in Table \ref{AxiomsForCKAL} for all $x,y,z,h\in \mathcal{T}_{SPR}$, where $x\leqq y$ means $x+y=y$.

\begin{center}
    \begin{table}
        \begin{tabular}{@{}ll@{}}
            \hline No. &Axiom\\
            $A1$ & $x+y=y+z$\\
            $A2$ & $x+(y+z)=(x+y)+z$\\
            $A3$ & $x+x=x$\\
            $A4$ & $(x+y)\cdot z=x\cdot z+y\cdot z$\\
            $A5$ & $x\cdot(y+z)=x\cdot y+x\cdot z$\\
            $A6$ & $x\cdot(y\cdot z)=(x\cdot y)\cdot z$\\
            $A7$ & $x+0=x$\\
            $A8$ & $0\cdot x=0$\\
            $A9$ & $x\cdot 0=0$\\
            $A10$ & $x\cdot 1=x$\\
            $A11$ & $1\cdot x=x$\\
            $P1$ & $x\parallel y=y\parallel x$\\
            $P2$ & $x\parallel(y\parallel z)=(x\parallel y)\parallel z$\\
            $P3$ & $(x+y)\parallel z=x\parallel z+y\parallel z$\\
            $P4$ & $x\parallel(y+z)=x\parallel y+x\parallel z$\\
            $P5$ & $(x\parallel y)\cdot (z\parallel h)\leqq(x\cdot z)\parallel(y\cdot h)$\\
            $P6$ & $x\parallel 0=0$\\
            $P7$ & $0\parallel x=0$\\
            $P8$ & $x\parallel 1=x$\\
            $P9$ & $1\parallel x=x$\\
            $A12$ & $1+x\cdot x^*=x^*$\\
            $A13$ & $1+x^*\cdot x=x^*$\\
            $A14$ & $x+y\cdot z\leqq z\Rightarrow y^*\cdot x\leqq z$\\
            $A15$ & $x+y\cdot z\leqq y\Rightarrow x\cdot z^*\leqq y$\\
            $P10$ & $1+x\parallel x^{\langle *\rangle}=x^{\langle *\rangle}$\\
            $P11$ & $1+x^{\langle *\rangle}\parallel x=x^{\langle *\rangle}$\\
            $P12$ & $x+y\parallel z\leqq z\Rightarrow y^{\langle *\rangle}\parallel x\leqq z$\\
            $P13$ & $x+y\parallel z\leqq y\Rightarrow x\parallel z^{\langle *\rangle}\leqq y$\\
        \end{tabular}
        \caption{Axioms of CKA modulo language equivalence}
        \label{AxiomsForCKAL}
    \end{table}
\end{center}

Since language equivalence is a congruence w.r.t. the operators of CKA, we can only check the soundness of each axiom in Table \ref{AxiomsForCKAL} according to the definition of semantics of spr-expressions. Then we can get the following soundness and completeness theorem \cite{CKA3} \cite{CKA4} \cite{CKA7}.

\begin{theorem}[Soundness and completeness of CKA]
For all $x,y\in\mathcal{T}_{SPR}$, $x= y$ if and only if $\sembrack{x}_{SPR}=\sembrack{y}_{SPR}$.
\end{theorem} 

\begin{theorem}
Let $x, y\in\mathcal{T}_{SPR}$. It is decidable whether $\sembrack{x}_{SPR}=\sembrack{y}_{SPR}$.
\end{theorem}

\begin{definition}
We define $\mathcal{F}_{SPR}$ as smallest subset of $\mathcal{T}_{SPR}$ satisfying the following rules:

$$\frac{}{1\in\mathcal{F}_{SPR}}\quad \frac{x\in\mathcal{F}_{SPR}\quad y\in\mathcal{T}_{SPR}}{x+y\in\mathcal{F}_{SPR}\quad y+x\in\mathcal{F}_{SPR}} \quad\frac{x\in\mathcal{T}_{SPR}}{x^*\in\mathcal{F}_{SPR}\quad x^{\langle *\rangle}\in\mathcal{F}_{SPR}}$$
$$\frac{x\in\mathcal{F}_{SPR}\quad y\in\mathcal{F}_{SPR}}{x\cdot y\in\mathcal{F}_{SPR}\quad x\parallel y\in\mathcal{F}_{SPR}}$$
\end{definition}

\subsection{Series-Parallel Rational Systems}\label{sprs}

\begin{definition}[Series-parallel rational system modulo language equivalence]
Let $Q$ be a finite set. A series-parallel rational system modulo language equivalence on $Q$, or called spr-system modulo language equivalence, is a pair $\mathcal{S}=\langle M,b\rangle$, where $M:Q^2\rightarrow\mathcal{T}_{SPR}$ and $b:Q\rightarrow\mathcal{T}_{SPR}$. Let $=$ be an CKA language equivalence on $\mathcal{T}_{SPR}(\Delta)$ with $\Sigma\subseteq\Delta$. We call $s:Q\rightarrow\mathcal{T}_{SPR}(\Delta)$ a $=$-solution to $\mathcal{S}$ if for $q\in Q$:

$$b(q)+\sum_{q'\in Q}M(q,q')\cdot s(q')\leqq s(q)$$

Lastly, $s$ is the least $=$-solution, if for every such solution $s'$ and every $q\in Q$, we have $s(q)\leqq s'(q)$.
\end{definition}

\begin{theorem}
Let $\mathcal{S}=\langle M,b\rangle$ be an spr-system on $Q$ modulo language equivalence. We can construct an $s:Q\rightarrow \mathcal{T}_{SPR}$ that, for any CKA equivalence $=$ on $\mathcal{T}_{SPR}(\Delta)$ with $\Sigma\subseteq\Delta$ and any $x\in\mathcal{T}_{SPR}$, the $Q$-vector $s:Q\rightarrow\mathcal{T}_{SPR}$ is the least $=$-solution to $\mathcal{S}$; we call such an $s$ the least solution to $\mathcal{S}$.
\end{theorem}

\newpage\section{Step Automata}\label{sa}

In this section, we introduce the so-called step automata which exactly accepts series-parallel rational (spr) language. We introduce the process from expressions to SAs in \cref{eta} and that from SAs to expressions in \cref{ate}.

\begin{definition}[Step automaton]
A step automaton (SA) is a tuple $A=(Q,F,\delta,\gamma)$ where:

\begin{enumerate}
  \item $Q$ is a finite set of states.
  \item $F\subseteq Q$ is the set of accepting states.
  \item $\delta:(Q\setminus F)\times\Sigma\rightarrow 2^Q$ is the sequential transition function which is the transition of traditional Kleene automata.
  \item $\gamma:(Q\setminus F)\times U\rightarrow 2^Q$ is the step transition function where $U\in \mathsf{SCP}(\Sigma)$ and $\mathsf{SCP}(\Sigma)$ is the series-parallel step of $\Sigma$, which is a step of actions.
\end{enumerate}
\end{definition}

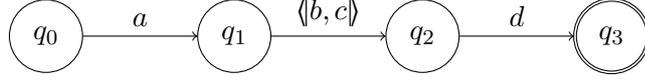
\begin{figure}
  \centering
  \begin{tikzpicture}[every node/.style={transform shape}]
    \node[state] (q0) {$q_0$};
    \node[state,right=15mm of q0] (q1) {$q_1$};
    \node[state,right=15mm of q1] (q2) {$q_2$};
    \node[state,accepting,right=15mm of q2] (q3) {$q_3$};

    \draw (q0) edge[->] node[above] {$a$} (q1);
    \draw (q1) edge[->] node[above] {$\semangle{b,c}$} (q2);
    \draw (q2) edge[->] node[above] {$d$} (q3);
  \end{tikzpicture}
  \caption{SA accepting $a \cdot (b \parallel c) \cdot d$.}\label{figure:example11}
\end{figure}

The SA accepting $a\cdot (b\parallel c)\cdot d$ is illustrated in Figure \ref{figure:example11}. 

\begin{definition}[Run relation]
Let $\mathsf{SCP}(\Sigma)$ be the series-parallel step of $\Sigma$, $a\in\Sigma$, $q,q'\in Q$ and $U\subseteq \mathsf{SCP}(\Sigma)$. We define the run relation $\xrightarrow[A]{}\subseteq Q\times \mathsf{SCP}(\Sigma)\times Q$ on a SA $A$ as the smallest relation satisfying:

\begin{enumerate}
  \item $q\xrightarrow[A]{1}q$ for all $q\in Q$.
  \item $q\xrightarrow[A]{a}q'$ if and only if $q'\in\delta(q,a)$.
  \item $q\xrightarrow[A]{U}q'$ if and only if $q'\in\gamma(q,U)$.
\end{enumerate}

Each $q\xrightarrow[A]{1}q$, $q\xrightarrow[A]{a}q'$, and $q\xrightarrow[A]{U}q'$ is called a unit run.
\end{definition}

\begin{definition}[Paths]
Let $A=(Q,F,\delta,\gamma)$ be an SA. We generalize the run relation $\rightarrow$ to paths $\xtworightarrow{}$, for $w\in \mathsf{SCP}(\Sigma)^*$ and $q_i,q_j\in Q$, then $q_i\xtworightarrow[A]{w}q_j$ denotes a path from $q_i$ to $q_j$ with label $w$, which can be derived as follows:

\begin{enumerate}
  \item For all $q\in Q$, it holds that $q\xtworightarrow[A]{1}q$.
  \item For all $q_i,q_j,q_k\in Q$, if $q_i\xrightarrow[A]{U}q_j$ and $q_j\xtworightarrow[A]{w}q_k$, then $q_i\xtworightarrow[A]{Uw}q_k$.
\end{enumerate}
\end{definition}

\begin{definition}[Language of SA]
The SA $A=(Q,A,\delta,\gamma)$ accepts the language by $q\in Q$, is the set $L_A(q)=\{w\in\mathsf{SCP}(\Sigma)^*:q\xtworightarrow[A]{w}q'\in F\}$. 
\end{definition}

It has already been proven that the so-called well-nested pomset automaton \cite{CKA7} just exactly accepts series-parallel rational (spr) language. In the following, we extend the related concepts and conclusions from \cite{CKA7} and prove that well-nested SA exactly accepts series-parallel rational (spr) languages. 

\begin{definition}[Support relation]
The support relation $\preceq$ of $A$ is the smallest preorder on $Q$ and for $q\in Q$:

$$\frac{a\in\Sigma\quad q'\in \delta(q,a)}{q'\preceq_{A}q}$$
$$\frac{U\in\mathsf{SCP}(\Sigma)\quad q'\in \gamma(q,U)}{q'\preceq_{A}q}$$

We call the strict support relation $\prec_A$ if $q'\preceq_A q$ and $q\npreceq_A q'$ then $q'\prec_A q$ holds.
\end{definition}

\begin{definition}[Support]
$Q'\subseteq Q$ is support-closed if for all $q\in Q'$ with $q'\preceq_A q$ then $q'\in Q'$. The support of $q\in Q$ denoted $\pi_A(q)$ is the smallest support-closed set containing $q$.
\end{definition}

\begin{definition}[Bounded]
If $\pi_A(q)$ is finite for all $q\in Q$ in $A$, then $A$ is called bounded.
\end{definition}

Parallel star allows an unbounded number of events to occur in parallel, we need the following concepts.

\begin{definition}[Pomset width]
The width of a finite pomset $U=[\mathbf{u}]\in\mathsf{Pom}$ is the size of maximum of the largest $\leq_{\mathbf{u}}$-antichain.
\end{definition}

\begin{definition}[Pomset depth]
The depth of $U\in \mathsf{SCP}$ denoted $d(U)$ is defined inductively as follows:

\begin{enumerate}
  \item $d(U)=0$ if $U$ is empty or primitive.
  \item $d(U)=1+\max_{1\leq i\leq n}d(U_i)$ if $U$ is sequential with sequential factorization $U_1,\cdots,U_n$.
  \item $d(U)=1+\max_{1\leq i\leq n}d(U_i)$ if $U$ is parallel with parallel factorization $\mset{U_1,\cdots,U_n}$.
\end{enumerate}
\end{definition}

\begin{definition}[Recursive states]
Let $A=(Q,F,\delta,\gamma)$ be a SA, $q\in Q$ is recursive if:

\begin{enumerate}
  \item For all $a\in\Sigma$, $q'\in \delta(q,a)$, then $q'\prec_{A}q$.
  \item For all $U\in\mathsf{SCP}(\Sigma)$, $q'\in \gamma(q,U)$, then $q'\prec_{A}q$.
\end{enumerate}
\end{definition}

\begin{definition}[Well-nestedness]
Let $A=(Q,F,\delta,\gamma)$ be a SA, $A$ is well-nested if every state is recursive.
\end{definition}

\begin{lemma}
Let SAs $A=(Q,F,\delta,\gamma)$ and $A[Q']=(Q',F\cap Q',\delta',\gamma')$, if $Q'$ is support-closed and $A$ is well-nested, then $A[Q']$ is well-nested.
\end{lemma}

\subsection{Expressions to Automata}\label{eta}

Given an spr-expression $x$, we show that how to obtain a well-nested and finite SA with some state accepting $\sembrack{x}_{SPR}$. Similarly to the process of pomset automaton accepting spr-expression, we firstly construct the so-called series-parallel rational syntactic SA.

\begin{definition}[Series-parallel rational syntactic step automaton]
Let $x\in\mathcal{T}_{SPR}$ and $S\subseteq\mathcal{T}_{SPR}$: (1)$x\star S=S$, if $x\in\mathcal{F}_{SPR}$; (2) $x\star S=\emptyset$, otherwise. We define the series-parallel rational syntactic SA as $A_{SPR}=(\mathcal{T}_{SPR},\mathcal{F}_{SPR},\delta_{SPR},\gamma_{SPR})$, where $\delta_{SPR}:\mathcal{T}_{SPR}\times\Sigma\rightarrow 2^{\mathcal{T}_{SPR}}$ is defined inductively as follows.

$$\delta_{SPR}(0,a)=\emptyset \quad \delta_{SPR}(1,a)=\emptyset \quad \delta_{SPR}(b,a)=\{1:a=b\}$$
$$\delta_{SPR}(x+y,a)=\delta_{SPR}(x,a)\cup\delta_{SPR}(y,a) \quad \delta_{SPR}(x\cdot y,a)=\delta_{SPR}(x,a)\fatsemi y\;\cup\; x\star\delta_{SPR}(y,a)$$
$$\delta_{SPR}(x^*,a)=\delta_{SPR}(x,a)\fatsemi x^* \quad \delta_{SPR}(x^{\langle *\rangle},a)=\emptyset \quad \delta_{SPR}(x\parallel y,a)=\emptyset$$

$\gamma_{SPR}:\mathcal{T}_{SPR}\times U\rightarrow 2^{\mathcal{T}_{SPR}}$ is defined inductively as follows.

$$\gamma_{SPR}(0,U)=\emptyset \quad \gamma_{SPR}(1,U)=\emptyset \quad \gamma_{SPR}(U,V)=\{1:U\sim V\}$$
$$\gamma_{SPR}(x+y,U)=\gamma_{SPR}(x,U)\cup\gamma_{SPR}(y,U) \quad \gamma_{SPR}(x\cdot y,U)=\gamma_{SPR}(x,U)\fatsemi y\;\cup\; x\star\gamma_{SPR}(y,U)$$
$$\gamma_{SPR}(x^*,U)=\gamma_{SPR}(x,U)\fatsemi x^* \quad \gamma_{SPR}(x^{\langle *\rangle},U)=\{1:U=\semangle{x,x^{\langle *\rangle}}\}\;\cup\;\{1\}\fatsemi\gamma_{SCPR}(x^{\langle *\rangle},U)$$
$$\gamma_{SPR}(x\parallel y,U)=\{1:U=\semangle{x,y}\}\;\cup\;\{U\}\fatsemi\gamma_{SCR}(x'\parallel y',U')\;\textrm{with}\;x\parallel y=U\cdot(x'\parallel y')$$
\end{definition}

\begin{lemma}
Let $x_1,x_2\in\mathcal{T}_{SPR}$ and $w\in\mathsf{SCP}(\Sigma)^*$. The following two conclusions are equivalent:

\begin{enumerate}
  \item There exists a $y\in\mathcal{F}_{SPR}$ such that $x_1+x_2\xtworightarrow[A_{SPR}]{w} y$.
  \item There exists a $y\in\mathcal{F}_{SPR}$ such that $x_1\xtworightarrow[A_{SPR}]{w} y$ or $x_2\xtworightarrow[A_{SPR}]{w} y$.
\end{enumerate}
\end{lemma}

\begin{lemma}
Let $x_1,x_2\in\mathcal{T}_{SPR}$, $w\in\mathsf{SCP}(\Sigma)^*$ and $\ell\in\mathbb{N}$. The following two conclusions are equivalent:

\begin{enumerate}
  \item There exists a $y\in\mathcal{F}_{SPR}$ such that $x_1\cdot x_2\xtworightarrow[A_{SPR}]{w} y$ of length $\ell$.
  \item $w=w_1\cdot w_2$, then there exist $y_1,y_2\in\mathcal{F}_{SPR}$ such that $x_1\xtworightarrow[A_{SPR}]{w_1} y_1$ or $x_2\xtworightarrow[A_{SPR}]{w_2} y_2$ of length at most $\ell$.
\end{enumerate}
\end{lemma}

\begin{lemma}
Let $x_1,x_2\in\mathcal{T}_{SPR}$, $w\in\mathsf{SCP}(\Sigma)^*$. The following two conclusions are equivalent:

\begin{enumerate}
  \item There exists a $y\in\mathcal{F}_{SPR}$ such that $x_1\parallel x_2\xtworightarrow[A_{SPR}]{w} y$.
  \item $w=w_1\parallel w_2$, then there exist $y_1,y_2\in\mathcal{F}_{SPR}$ such that $x_1\xtworightarrow[A_{SPR}]{w_1} y_1$ or $x_2\xtworightarrow[A_{SPR}]{w_2} y_2$.
\end{enumerate}
\end{lemma}

\begin{lemma}
Let $x\in\mathcal{T}_{SPR}$, $w\in\mathsf{SCP}(\Sigma)^*$. The following two conclusions are equivalent:

\begin{enumerate}
  \item There exists a $y\in\mathcal{F}_{SPR}$ such that $x^*\xtworightarrow[A_{SPR}]{w} y$.
  \item $w=w_1\cdots w_n$, then there exist $y_i\in\mathcal{F}_{SPR}$ such that $x\xtworightarrow[A_{SPR}]{w_i} y_i$ for $1\leq i\leq n$.
\end{enumerate}
\end{lemma}

\begin{lemma}
Let $x\in\mathcal{T}_{SPR}$, $w\in\mathsf{SCP}(\Sigma)^*$. The following two conclusions are equivalent:

\begin{enumerate}
  \item There exists a $y\in\mathcal{F}_{SPR}$ such that $x^{\langle *\rangle}\xtworightarrow[A_{SPR}]{w} y$.
  \item $w=w_1\parallel\cdots\parallel w_n$, then there exist $y_i\in\mathcal{F}_{SPR}$ such that $x\xtworightarrow[A_{SPR}]{w_i} y_i$ for $1\leq i\leq n$.
\end{enumerate}
\end{lemma}

\begin{lemma}
Let $x,y\in\mathcal{T}_{SPR}$, then the following hold:

\begin{enumerate}
  \item $L_{SPR}(x+y)=L_{SPR}(x)+L_{SPR}(y)$.
  \item $L_{SPR}(x\cdot y)=L_{SPR}(x)\cdot L_{SPR}(y)$.
  \item $L_{SPR}(x^*)=L_{SPR}(x)^*$.
  \item $L_{SPR}(x^{\langle *\rangle})=L_{SPR}(x)^{\langle *\rangle}$.
  \item $L_{SPR}(x\parallel y)=L_{SPR}(x)\parallel L_{SPR}(y)$.
\end{enumerate}
\end{lemma}

\begin{lemma}
For all $x\in\mathcal{T}_{SPR}$, it holds that $L_{SPR}(x)=\sembrack{x}_{SPR}$.
\end{lemma}

\begin{definition}[$\parallel$-depth]
The $\parallel$-depth of $x\in\mathcal{T}_{SPR}$ denoted $d_{\parallel}(x)$ is defined inductively on the structure of $x$ as follows.

$$d_{\parallel}(0)=0\quad d_{\parallel}(1)=0\quad d_{\parallel}(a)=0\quad d_{\parallel}(U)=0$$
$$d_{\parallel}(x\cdot y)=\max(d_{\parallel}(x),d_{\parallel}(y))\quad d_{\parallel}(x+ y)=\max(d_{\parallel}(x),d_{\parallel}(y))\quad d_{\parallel}(x^*)=d_{\parallel}(x)$$ 
$$d_{\parallel}(x\parallel y)=\max(d_{\parallel}(x),d_{\parallel}(y))+1\quad d_{\parallel}(x^{\langle *\rangle})=d_{\parallel}(x)$$
\end{definition}

\begin{definition}[$\langle *\rangle$-depth]
We define the $\langle *\rangle$-depth of $x\in\mathcal{T}_{SPR}$ denoted $d_{\langle *\rangle}(x)$ is defined inductively on the structure of $x$ as follows.

$$d_{\langle *\rangle}(0)=0\quad d_{\langle *\rangle}(1)=0\quad d_{\langle *\rangle}(a)=0\quad d_{\langle *\rangle}(U)=0$$
$$d_{\langle *\rangle}(x\cdot y)=\max(d_{\langle *\rangle}(x),d_{\langle *\rangle}(y))\quad d_{\langle *\rangle}(x+ y)=\max(d_{\langle *\rangle}(x),d_{\langle *\rangle}(y))\quad d_{\langle *\rangle}(x^*)=d_{\langle *\rangle}(x)$$ 
$$d_{\langle *\rangle}(x\parallel y)=\max(d_{\langle *\rangle}(x),d_{\langle *\rangle}(y))\quad d_{\langle *\rangle}(x^{\langle *\rangle})=d_{\langle *\rangle}(x)+1$$
\end{definition}

\begin{lemma}
Let $x,y\in\mathcal{T}_{SPR}$, if $x\preceq_{SPR} y$, then $d_{\parallel}(x)\leq d_{\parallel}(y)$ and $d_{\langle *\rangle}(x)\leq d_{\langle *\rangle}(y)$.
\end{lemma}

\begin{lemma}
Let $x,y\in\mathcal{T}_{SPR}$, if $x\preceq_{SPR} y^{\langle *\rangle}$ and $d_{\langle *\rangle}(x)=d_{\langle *\rangle}(y^{\langle *\rangle})$, then $x=y^{\langle *\rangle}$.
\end{lemma}

\begin{lemma}
Every $x\in\mathcal{T}_{SPR}$ is recursive in $A_{SPR}$, and the syntactic SA is well-nested.
\end{lemma}

\begin{definition}
Let $x_1,x_2\in\mathcal{T}_{SPR}$, $R:\mathcal{T}_{SPR}\rightarrow 2^{\mathcal{T}_{SPR}}$ is defined inductively as follows:

$$R(0)=\{0\}\quad R(1)=\{1\} \quad R(a)=\{a,1\}\quad R(U)=\{U,1\}$$
$$R(x_1+x_2)=R(x_1)\cup R(x_2)\quad R(x_1\cdot x_2)=R(x_1)\fatsemi x_2\;\cup\; R(x_1)\cup R(x_2)$$
$$R(x_1^*)=R(x_1)\fatsemi x_1^*\;\cup\; R(x_1)\cup\{x_1^*\}\quad R(x_1^{\langle *\rangle})=R(x_1)\cup\{x_1^{\langle *\rangle},1\} \quad R(x_1\parallel x_2)=R(x_1)\cup R(x_2)\cup\{x_1\parallel x_2,1\}$$
\end{definition}

\begin{lemma}
For every $x\in\mathcal{T}_{SPR}$, they hold that:

\begin{enumerate}
  \item $x\in R(x)$.
  \item $R(x)$ is support-closed.
  \item The syntactic SA is bounded.
\end{enumerate}
\end{lemma}

\begin{theorem}[Expressions to automata]\label{etoa2}
For every $x\in\mathcal{T}_{SPR}$, we can obtain a well-nested and finite SA $A$ with a state $q$ such that $L_{A}(q)=\sembrack{x}_{SPR}$.
\end{theorem}

\subsection{Automata to Expressions}\label{ate}

In this section, we show that the language accepted by a state in any well-nested and finite automaton can be implemented by a series-parallel rational expression.

\begin{definition}[Solution of a SA]
Let $A=(Q,F,\delta,\gamma)$ be a SA, and let $=$ be an CKA language congruence on $\mathcal{T}_{SPR}(\Delta)$ with $\Sigma\subseteq\Delta$. We say that $s:Q\rightarrow\mathcal{T}_{SPR}(\Delta)$ is an $=$-solution to $A$, if for every $q\in Q$:

\begin{center}
$[q\in \mathcal{F}_{SPR}]+\sum_{q'\in\delta(q,a)}a\cdot s(q')+\sum_{q'\in\gamma(q,U)}U\cdot s(q')\leqq s(q)$
\end{center}

Also, $s$ is a least $=$-solution to $A$ if for every $=$-solution $s'$ it holds that $s(q)\leqq s'(q)$ for all $q\in Q$. We call $s:Q\rightarrow\mathcal{T}_{SPR}$ the least solution to $A$ if it is the least $=$-solution for any CKA language congruence $=$.
\end{definition}

\begin{lemma}
Let $A=(Q,F,\delta,\gamma)$ be a step automaton. If $s:Q\rightarrow\mathcal{T}_{SPR}$ is the least solution to $A$, then it holds that $L_A(q)=\sembrack{s(q)}_{SPR}$ for $q\in Q$.
\end{lemma}

\begin{lemma}
Let $A$ be a well-nested and finite SA, then we can construct the least solution to $A$.
\end{lemma}

\begin{theorem}[Automata to expressions]
If $A=(Q,F,\delta,\gamma)$ is a well-nested and finite SA, then we can construct for every $q\in Q$ a series-parallel rational expression $x\in\mathcal{T}_{SPR}$ such that $L_{A}(q)=\sembrack{x}_{SPR}$.
\end{theorem}

\begin{corollary}[Kleene theorem for series-parallel rational language]
Let $L\subseteq\mathsf{SCP}$, then $L$ is series-parallel rational if and only if it is accepted by a finite and well-nested step automaton.
\end{corollary} 
\newpage\section{Step Turing Machine}\label{stm} 

In this section, we introduce step Turing machine which is a step automaton equipped with a kind of infinite memories called planar step tape. We introduce the definition of a step Turing machine in \cref{stm2} and its computability in \cref{ctt}.

\subsection{Step Turing Machine}\label{stm2}

Firstly, we give the definition of step Turing machine.

\begin{figure}[!htbp]
 \centering
 \includegraphics{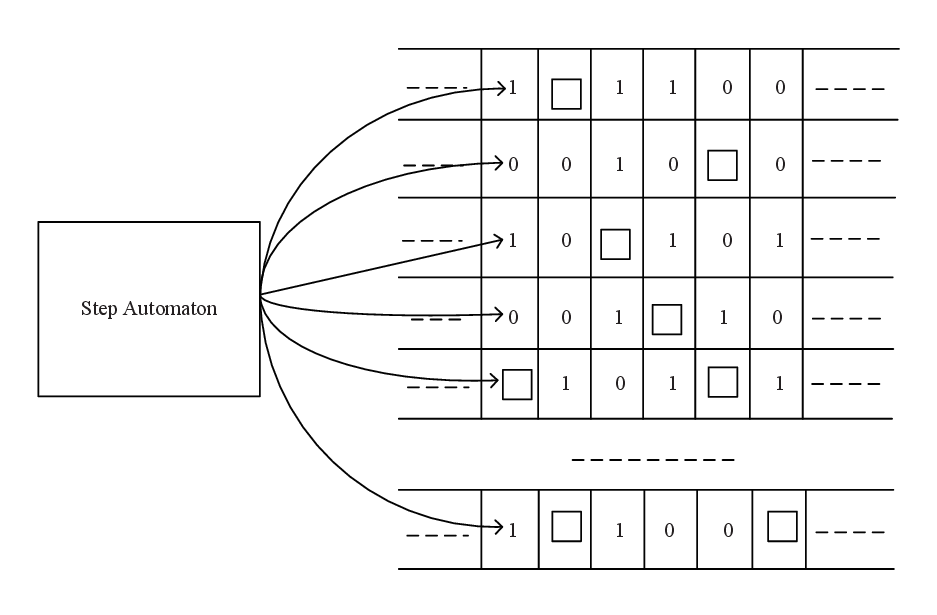}
 \caption{Step Turing machine}
 \label{steptm}
\end{figure} 

\begin{definition}[Step Turing machine]
A step Turing machine (STM) is defined as an octuple $M=(Q,q_0,F,\Sigma,\Gamma,\delta,\gamma,\eta)$ where

\begin{itemize}
  \item $Q$ is a finite set of control states.
  \item $q_0\in Q$ is the initial state.
  \item $F\subseteq Q$ is the set of final states.
  \item $\Sigma$ is a finite alphabet including special actions 0, 1.
  \item $\Gamma$ is the set of tape symbols consisting of $0$, $1$ and the blank symbol $\square$.
  \item $\delta: (Q\setminus F)\times 1\times \Gamma\rightarrow Q\times\Gamma^{\langle *\rangle^*}_{\ell r}\times (D=\{R\})$ is a finite set of input reading transitions, where $D=\{R\}$ is the read head's moving direction: only right. For $(q_i,1,\Gamma)\rightarrow (q_j,\Gamma^{\langle *\rangle^*}_{\ell r},R)$, we write it as $q_i\xrightarrow{1,\Gamma,\Gamma^{\langle *\rangle^*}_{\ell r},R}q_j$, where $q_i,q_j\in Q$, $\ell$ is the current location of read/wirte head of the planar step tape and $r$ is the $r$-th cell from top to bottom. It means that the STM is in state $q_i$ and reading symbol $\Gamma$ from the input tape, will execute an empty action $1$, write it into $\ell r$-th cell of the planar step tape, move one step right on the input tape, and therefore evolve into state $q_j$.
  \item $\gamma: (Q\setminus F)\times 1\times\Gamma^{\langle *\rangle^*}_{\ell r}\rightarrow Q\times \Gamma\times (D=\{R\})$ is a finite set of output writing transitions, where $D=\{R\}$ is the write head's moving direction: only right. For $(q_i,1,\Gamma^{\langle *\rangle^*}_{\ell r})\rightarrow (q_j,\Gamma,R)$, we write it as $q_i\xrightarrow{1,\Gamma^{\langle *\rangle^*}_{\ell r},\Gamma,R}q_j$, where $q_i,q_j\in Q$, $\ell$ is the current location of read/wirte head of the planar step tape and $r$ is the $r$-th cell from top to bottom. It means that the STM is in state $q_i$ and reading the symbol $\Gamma^{\langle *\rangle^*}_{\ell r}$ from $\ell r$-th cell of the planar step tape, writing it to the output tape, will execute an empty action $1$, move one step right on the output tape, and therefore evolve into state $q_j$.
  \item $\eta: (Q\setminus F)\times U\times (\Gamma\cup\{\epsilon\})^{\langle k\rangle}\rightarrow Q\times (\Gamma\cup\{\epsilon\})^{\langle k\rangle}\times (D=\{L,R\})$ is a finite set of step transitions, where $U\in \mathsf{SCP}(\Sigma)$ and $\mathsf{SCP}(\Sigma)$ is the series-parallel step of $\Sigma$, $k\in\mathbb{N}$,  $D=\{L,R\}$ is the read/write heads' moving direction: left or right, and if $\epsilon$ is read, it means that we are looking at an empty part of the tape; if $\epsilon$ is written, then a symbol on the tape will be erased. For $(q_i,U,(\Gamma\cup\{\epsilon\})^{\langle k\rangle})\rightarrow (q_j, (\Gamma\cup\{\epsilon\})^{\langle k\rangle},D)$, we write it as $q_i\xrightarrow{U,(\Gamma\cup\{\epsilon\})^{\langle k\rangle},(\Gamma\cup\{\epsilon\})^{\langle k\rangle},D}q_j$, where $q_i,q_j\in Q$. It means that the STM is in state $q_i$ and reading symbols $(\Gamma\cup\{\epsilon\})^{\langle k\rangle}$ on the planar tape in parallel, will execute a step of actions $U$, change the symbols on the planar tape to new symbols $(\Gamma\cup\{\epsilon\})^{\langle k\rangle}$, move one step left if $D=L$ and right if $D=R$, and therefore evolve into state $q_j$. Note that, when $U=1$, the current contents of the planar step tape remain unchanged, and then the heads move left or right. 
\end{itemize}

As \cref{steptm} illustrates, an STM has a step automaton providing finite state control, and a planar tape within which exists finite classical tapes and each classical tape is with a read/write head and has infinite cells. Each cell may contain a tape symbol either being $0$, or $1$, or $\square$. The vertical dimension of the planar tape has an upper bound and a lower bound. The first classical tape is the read-only input tape and the second classical tape is the write-only output tape. The whole planar step tape move left/right, and read/write contents from/into the cells within a step in parallel.
\end{definition}

The STM is off-line, i.e., input to an STM is written on the input tape with an element of $\{0,1\}^*$ and the input tape head is on the left most nonblank cell. The output of an STM is the contents of the output tape when the STM halts and ensures that it is an element of $\{0,1\}^*$.

The configurations of an STM are determined by the contents of its three kinds of tapes and the locations of the corresponding read/write heads.

\begin{definition}[Configurations of an STM]
The current configuration of an STM consists of the contents and the current location of read head of the input tape, the contents and the current location of write head of the output tape, and the contents and the current location of read/write heads of the planar step tape. We use a bar over a symbol of the contents to denote the current location of the tape head, for example, $\square1\bar{0}1\square$ is a configuration of a classical Turing machine.

\begin{itemize}
  \item For the input tape with input contents $101$, the possible configurations are $\bar{\square}101\square$, $\square\bar{1}01\square$, $\square1\bar{0}1\square$, $\square10\bar{1}\square$ and $\square101\bar{\square}$.
  \item For the output tape with output contents $101$, $\square101\bar{\square}$ is the only possible configuration.
  \item For the planar step tape with intermediate contents:
  $$
  \left[\begin{array}{ccc}
     0 & 1 & 0\\
     1 & 1 & 0\\
     1 & 0 & 1\\
     0 &   & 0\\
       &   & 1
  \end{array}\right]
  $$
  the possible configurations are:
  \begin{small}
  $$
  \left[\begin{array}{ccccc}
     \bar{\square} & 0 & 1 & 0 & \square\\
     \bar{\square} & 1 & 1 & 0 & \square\\
     \bar{\square} & 1 & 0 & 1 & \square\\
     \bar{\square} & 0 &   & 0 & \square\\
     \bar{\square} &   &   & 1 & \square
  \end{array}\right]
  \left[\begin{array}{ccccc}
     \square & \bar{0} & 1 & 0 & \square\\
     \square & \bar{1} & 1 & 0 & \square\\
     \square & \bar{1} & 0 & 1 & \square\\
     \square & \bar{0} &   & 0 & \square\\
     \square &   &   & 1 & \square
  \end{array}\right]
  \left[\begin{array}{ccccc}
     \square & 0 & \bar{1} & 0 & \square\\
     \square & 1 & \bar{1} & 0 & \square\\
     \square & 1 & \bar{0} & 1 & \square\\
     \square & 0 &   & 0 & \square\\
     \square &   &   & 1 & \square
  \end{array}\right]
  \left[\begin{array}{ccccc}
     \square & 0 & 1 & \bar{0} & \square\\
     \square & 1 & 1 & \bar{0} & \square\\
     \square & 1 & 0 & \bar{1} & \square\\
     \square & 0 &   & \bar{0} & \square\\
     \square &   &   & \bar{1} & \square
  \end{array}\right]
  \left[\begin{array}{ccccc}
     \square & 0 & 1 & 0 & \bar{\square}\\
     \square & 1 & 1 & 0 & \bar{\square}\\
     \square & 1 & 0 & 1 & \bar{\square}\\
     \square & 0 &   & 0 & \bar{\square}\\
     \square &   &   & 1 & \bar{\square}
  \end{array}\right]
  $$
  \end{small}
\end{itemize}
\end{definition}

\begin{definition}[States of an STM]
Let $M=(Q,q_0,F,\Sigma,\Gamma,\delta,\gamma,\eta)$ be an STM. The set of states is determined by the finite control states and the configurations, i.e., $\{(q,\bar{\square})|q\in Q,\square\\\textrm{is the beginning of the input tape}\}\cup\{(q,\square x\square,\bar{ })|q\in Q,x\in\Gamma^*\textrm{ is the input}\}\cup\{(q,\bar{\square})|q\in Q,\square\\\textrm{is the beginning of the output tape}\}\cup\{(q,\square x\bar{\square})|q\in Q,x\in\Gamma^*\textrm{ is the output}\}\cup\{(q,\square^{\langle *\rangle})|q\in Q,\square^{\langle *\rangle} \\\textrm{is the beginning of the planar step tape}\}\cup\{(q,\square^{\langle *\rangle}x\square^{\langle *\rangle},\bar{ })|q\in Q,x\in\Gamma^{\langle *\rangle^*}\}$, where $\bar{ }$ denotes the location of read/write head, on one of the elements of $\square x\square$ or $\square^{\langle *\rangle}x\square^{\langle *\rangle}$.

The initial state is $(q_0,\bar{\square},\bar{\square},\overline{\square^{\langle *\rangle}})$, and if $(q,\bar{\square},\bar{\square},\overline{\square^{\langle *\rangle}})\in F$, then $q\in F$.
\end{definition}

\begin{definition}[Transition system of an STM]
The transition system of an STM includes three parts: the one of reading inputs, the one of writing outputs and the one of intermediate steps.

\begin{itemize}
  \item Transition system of reading inputs:
  
  $(q_i,\square x\bar{d}fy\square)\xrightarrow{1} (q_j,\square xd\bar{f}y\square,\square^{\langle *\rangle}\Gamma^{\langle *\rangle^*}\square^{\langle *\rangle})$, if and only if $q_i\xrightarrow{1,d,\Gamma^{\langle *\rangle^*}_{\ell r}=d,R}q_j$, where $d,f\in\Gamma, x,y\in\Gamma^*$, $\Gamma^{\langle *\rangle^*}_{\ell r}$ is the $\ell r$-th cell of the planar step tape. 
  \item Transition system of writing outputs:
  
  $(q_i,1,\square x\bar{\square},\square^{\langle *\rangle}\Gamma^{\langle *\rangle^*}\square^{\langle *\rangle})\xrightarrow{1} (q_j,\square x \Gamma^{\langle *\rangle^*}_{\ell r}\bar{\square})$, if and only if $q_i\xrightarrow{1,\Gamma^{\langle *\rangle^*}_{\ell r},\Gamma^{\langle *\rangle^*}_{\ell r},R}q_j$, where $x\in\Gamma^*$, $\Gamma^{\langle *\rangle^*}_{\ell r}$ is the $\ell r$-th cell of the planar step tape.
  \item Transition system of the intermediate steps:
  
  $(q_i,\square^{\langle *\rangle}x\bar{d}\square^{\langle *\rangle})\xrightarrow{U}(q_j,\square^{\langle *\rangle}xe\overline{\square^{\langle *\rangle}})$ if and only if $q_i\xrightarrow{U,d,e,R}q_j$, where $d,e\in\Gamma^{\langle k\rangle},x\in\Gamma^{\langle *\rangle^*}$.
  
  $(q_i,\square^{\langle *\rangle}x\bar{d}fy\square^{\langle *\rangle})\xrightarrow{U}(q_j,\square^{\langle *\rangle}xe\bar{f}y\square^{\langle *\rangle})$ if and only if $q_i\xrightarrow{U,d,e,R}q_j$, where $d,e,f\in\Gamma^{\langle k\rangle},x,y\in\Gamma^{\langle *\rangle^*}$.
  
  $(q_i,\square^{\langle *\rangle}\bar{d}x\square^{\langle *\rangle})\xrightarrow{U}(q_j,\overline{\square^{\langle *\rangle}}ex\square^{\langle *\rangle})$ if and only if $q_i\xrightarrow{U,d,e,L}q_j$, where $d,e\in\Gamma^{\langle k\rangle},x\in\Gamma^{\langle *\rangle^*}$.
  
  $(q_i,\square^{\langle *\rangle}xf\bar{d}y\square^{\langle *\rangle})\xrightarrow{U}(q_j,\square^{\langle *\rangle}x\bar{f}ey\square^{\langle *\rangle})$ if and only if $q_i\xrightarrow{U,d,e,L}q_j$, where $d,e,f\in\Gamma^{\langle k\rangle},x,y\in\Gamma^{\langle *\rangle^*}$.
  
  $(q_i,\square^{\langle *\rangle}\bar{d}\square^{\langle *\rangle})\xrightarrow{U}(q_j,\square^{\langle *\rangle}\overline{\square^{\langle *\rangle}})$ if and only if $q_i\xrightarrow{U,d,\epsilon,R}q_j$, where $d\in\Gamma^{\langle k\rangle}$.
  
  $(q_i,\square^{\langle *\rangle}x\bar{d}\square^{\langle *\rangle})\xrightarrow{U}(q_j,\square^{\langle *\rangle}x\overline{\square^{\langle *\rangle}})$ if and only if $q_i\xrightarrow{U,d,\epsilon,R}q_j$, where $d\in\Gamma^{\langle k\rangle},x\in\Gamma^{\langle *\rangle^*}$.
  
  $(q_i,\square^{\langle *\rangle}\bar{d}fx\square^{\langle *\rangle})\xrightarrow{U}(q_j,\square^{\langle *\rangle}\bar{f}x\square^{\langle *\rangle})$ if and only if $q_i\xrightarrow{U,d,\epsilon,R}q_j$, where $d,f\in\Gamma^{\langle k\rangle},x\in\Gamma^{\langle *\rangle^*}$.
  
  $(q_i,\square^{\langle *\rangle}\bar{d}\square^{\langle *\rangle})\xrightarrow{U}(q_j,\overline{\square^{\langle *\rangle}})$ if and only if $q_i\xrightarrow{U,d,\epsilon,L}q_j$, where $d\in\Gamma^{\langle k\rangle}$.
  
  $(q_i,\square^{\langle *\rangle}\bar{d}x\square^{\langle *\rangle})\xrightarrow{U}(q_j,\overline{\square^{\langle *\rangle}}x\square^{\langle *\rangle})$ if and only if $q_i\xrightarrow{U,d,\epsilon,L}q_j$, where $d\in\Gamma^{\langle k\rangle},x\in\Gamma^{\langle *\rangle^*}$.
  
  $(q_i,\square^{\langle *\rangle}xf\bar{d}\square^{\langle *\rangle})\xrightarrow{U}(q_j,\square^{\langle *\rangle}x\bar{f}\square^{\langle *\rangle})$ if and only if $q_i\xrightarrow{U,d,\epsilon,L}q_j$, where $d,f\in\Gamma^{\langle k\rangle},x\in\Gamma^{\langle *\rangle^*}$.
  
  $(q_i,\overline{\square^{\langle *\rangle}})\xrightarrow{U}(q_j,\square^{\langle *\rangle}d\overline{\square^{\langle *\rangle}})$ if and only if $q_i\xrightarrow{U,\epsilon,d,R}q_j$, where $d\in\Gamma^{\langle k\rangle}$.
  
  $(q_i,\overline{\square^{\langle *\rangle}}fx\square^{\langle *\rangle})\xrightarrow{U}(q_j,\square^{\langle *\rangle}d\bar{f}x\square^{\langle *\rangle})$ if and only if $q_i\xrightarrow{U,\epsilon,d,R}q_j$, where $d\in\Gamma^{\langle k\rangle},x\in\Gamma^{\langle *\rangle^*}$.
  
  $(q_i,\overline{\square^{\langle *\rangle}})\xrightarrow{U}(q_j,\overline{\square^{\langle *\rangle}}d\square^{\langle *\rangle})$ if and only if $q_i\xrightarrow{U,\epsilon,d,L}q_j$, where $d\in\Gamma^{\langle k\rangle}$.
  
  $(q_i,\square^{\langle *\rangle}xf\overline{\square^{\langle *\rangle}})\xrightarrow{U}(q_j,\square^{\langle *\rangle}x\bar{f}d\square^{\langle *\rangle})$ if and only if $q_i\xrightarrow{U,\epsilon,d,L}q_j$, where $d,f\in\Gamma^{\langle k\rangle},x\in\Gamma^{\langle *\rangle^*}$.
  
  $(q_i,\overline{\square^{\langle *\rangle}})\xrightarrow{U}(q_j,\overline{\square^{\langle *\rangle}})$ if and only if $q_i\xrightarrow{U,\epsilon,\epsilon,R}q_j$.
  
  $(q_i,\overline{\square^{\langle *\rangle}}fx\square^{\langle *\rangle})\xrightarrow{U}(q_j,\square^{\langle *\rangle}\bar{f}x\square^{\langle *\rangle})$ if and only if $q_i\xrightarrow{U,\epsilon,\epsilon,R}q_j$, where $d\in\Gamma^{\langle k\rangle}$.
  
  $(q_i,\overline{\square^{\langle *\rangle}})\xrightarrow{U}(q_j,\overline{\square^{\langle *\rangle}})$ if and only if $q_i\xrightarrow{U,\epsilon,\epsilon,L}q_j$.
  
  $(q_i,\square^{\langle *\rangle}xf\overline{\square^{\langle *\rangle}})\xrightarrow{U}(q_j,\square^{\langle *\rangle}x\bar{f}\square^{\langle *\rangle})$ if and only if $q_i\xrightarrow{U,\epsilon,\epsilon,L}q_j$, where $f\in\Gamma^{\langle k\rangle}, x\in\Gamma^{\langle *\rangle^*}$.
\end{itemize}
\end{definition}

Similarly to the definition of path for step automata, we can define the path $\xtworightarrow{}$ of transition $\rightarrow$.

\begin{definition}[Language accepted by an STM]
Let $M=(Q,q_0,F,\Sigma,\Gamma,\delta,\gamma,\eta)$ be an STM. The language accepted by $M$, is the set $L_M=\{w\in\mathsf{SCP}(\Sigma)^*:(q_0,\bar{\square},\bar{\square},\overline{\square^{\langle *\rangle^*}})\xtworightarrow{w}(q',\bar{\square},\bar{\square},\overline{\square^{\langle *\rangle^*}}),q'\in F\}$. 
\end{definition}

\subsection{Church-Turing Thesis}\label{ctt}

\begin{definition}[Computation of an STM]
An atomic action $a\in\Sigma$ is the one that an STM can compute individually, and correspondingly, a step contains some atomic actions without partial orders between them pairwise. We say the computation of an STM, if the STM do computations step by step from the initial state to one of the final states. And we say that the STM computes the function $f$ on a domain of data $\Gamma^*$ if for all inputs $w\in\Gamma^*$, it has the outputs $f(w)$.
\end{definition}

\begin{definition}[Computability]
 A problem is computable if it can be computed by a classical Turing machine. An algorithm for a function is a classical Turing machine computing this function.
\end{definition} 

\begin{theorem}
The class of computable languages and functions defined by an STM is the same as the class of computable languages and functions defined by a classical Turing machine.
\end{theorem} 
\bibliographystyle{elsarticle-num}
\newpage\bibliography{Refs-CFMAI}

\end{document}